\documentclass[aps,pra,amsmath,amssymb,amsfonts,showpacs,superscriptaddress,floatfix,twocolumn]{revtex4}

\usepackage{amssymb}

\def\ket#1{\vert#1\rangle}
\def\ketbra#1{\vert#1\rangle\langle#1\vert}

\def\Q{Q}
\newcommand{\e}{{\mathrm e}}

\begin{document}

\title{Interacting boson problems are QMA-hard}
\author{Tzu-Chieh Wei}
\affiliation{Institute for Quantum Computing, University of Waterloo,
Waterloo, Ontario, Canada} \affiliation{Department of Physics and Astronomy,
University of Waterloo, Waterloo, Ontario, Canada}
\author{Michele Mosca}
\affiliation{Institute for Quantum Computing, University of Waterloo,
Waterloo, Ontario, Canada} \affiliation{Department of Combinatorics and
Optimization, University of Waterloo, Waterloo, Ontario, Canada}
\affiliation{St. Jerome's University, Waterloo, Ontario, Canada}
\affiliation{Perimeter Institute for Theoretical Physics, Waterloo, Ontario,
Canada}
 \author{Ashwin Nayak}
\affiliation{Department of Combinatorics and Optimization, University of
Waterloo, Waterloo, Ontario, Canada}
\affiliation{Institute for Quantum Computing, University of Waterloo,
Waterloo, Ontario, Canada}
\affiliation{Perimeter Institute for Theoretical Physics,
Waterloo, Ontario, Canada}

\date{May 20, 2009}

\begin{abstract}
  Computing the ground-state energy of interacting electron (fermion)
  problems has recently been shown to be hard for QMA, a quantum
  analogue of the complexity class NP. Fermionic problems are usually
  hard, a phenomenon widely attributed to the so-called sign problem
  occurring in Quantum Monte Carlo simulations. The corresponding
  bosonic problems are, according to conventional wisdom,
  tractable. Here, we discuss the complexity of interacting boson
  problems and show that they are also QMA-hard. In addition, we
  show that the bosonic version of the so-called $N$-representability
  problem is QMA-complete, as hard as its fermionic version. As a
  consequence, these problems are unlikely to have efficient quantum
  algorithms.
\end{abstract}

\maketitle

Many important model Hamiltonians in physics, such as the Hubbard model (both
fermionic and bosonic) and those for superconductivity and the quantum Hall
effect, involve at most two-body interactions~\cite{AshcroftMermin}. The
ground-state wavefunction and energy of these Hamiltonians play a key role in
understanding these fascinating phenomena. In some of these phenomena,
electrons are the major players. Problems involving fermionic particles such
as electrons often seem to be computationally more difficult than those with
their bosonic counterparts. This intractability is often attributed to the
so-called ``fermion sign problem'', occurring in Quantum Monte Carlo
simulations~\cite{TroyerWiese}. On the other hand, bosonic problems do not
suffer from the sign problem~\cite{Ceperley} and are thus regarded as
tractable.

Indeed, Schuch and Verstraete and Liu, Christandl and Verstraete have recently
shown that computing the ground-state energy of general interacting electrons
is QMA-hard~\cite{SchuchVerstraete,LiuChristandlVerstraete07}. The complexity
class QMA (Quantum Merlin-Arthur) is a generalization of the class NP
(nondeterministic polynomial time) to the quantum realm. It was introduced by
Kitaev~\cite{Kitaev} in the study of the so-called local Hamiltonian problem,
where, roughly speaking, the goal is to determine the ground-state energy of a
spin Hamiltonian involving only few-body interaction terms. QMA-hard problems
are regarded as difficult, unlikely to be solved efficiently even by a quantum
computer. However, a quantum computer, if given the solution to any problem in
QMA, along with a suitable ``certificate'' or ``witness state'', can
efficiently verify whether the solution is correct or not. In fact, as a
result of a series of works~\cite{Kempe1,Kempe2,Terhal}, even for
nearest-neighbor two-body interactions among spin-$1/2$ particles on a
two-dimensional lattice, the local Hamiltonian problem is QMA-complete. With
higher magnitude of spins in one dimension, the local Hamiltonian problem can
be QMA-complete as well~\cite{AGIK09}. Understanding and classifying the
complexity of physical models and investigating hard problems using
statistical mechanical tools have become important research
endeavors~\cite{Kitaev,SchuchVerstraete,Kempe1,Kempe2,Terhal,AGIK09,Kay,Coppersmith,Bravyi,Barahona,LiuChristandlVerstraete07,QSAT,NagajMozes07},
as a result of interplay between physics, mathematics and computer science.

The fact that interacting fermion problems are hard motivates us to
investigate the corresponding bosonic problems. The complexity of bosonic
problems seem less explored. Could bosonic problems be so hard as to be
intractable? We show that generic nearest-neighbor two-body interacting boson
problems are indeed QMA-hard. Inspired by the recent work of Liu, Christandl,
and Verstraete~\cite{LiuChristandlVerstraete07} that shows QMA-completeness of
the (fermionic) $N$-representability problem~\cite{Coleman,Coulson}, we study
its bosonic version. In an $N$-representability problem one is given a
two-particle reduced density matrix $\rho$ and needs to decide whether there
exists a global $N$-body state $\sigma$ that is consistent with $\rho$. The
$N$-representability problem has been vastly explored in quantum
chemistry~\cite{QChemistry}, and its solution would enable efficient solution
of ground-state energy for generic two-body interacting fermionic systems. We
show that the bosonic $N$-representability problem is also QMA-complete.
Similarly, we show that the bosonic $N$-representability problem given only
diagonal elements is NP-hard.
\medskip

\noindent {\it QMA-hardness of interacting boson problems.\/} Let us
consider boson creation and annihilation operators $a_j^\dagger$,
$a_j$ for the $j$'th site or mode, and denote by $\ket{\Omega}$ the
vacuum state without any bosons. The creation and annihilation
operators obey the following commutation
relations~\cite{NegeleOrland}:
\begin{equation}
[a_i,a_j] = 0 = [a_i^\dagger,a_j^\dagger], \hspace{1cm}
[a_i,a_j^\dagger] = \delta_{ij}.
\end{equation}
The use of these operators preserves the symmetry of the bosonic wavefunctions
under permutations, and any $N$-boson wavefunction can be represented in the
following second-quantization formalism:
\begin{equation}
\label{eqn:psiN} |\psi\rangle = \sum_{ i_1+...+i_m=N} c_{i_1,...,i_m}
(a_1^\dagger)^{i_1}...(a_m^\dagger)^{i_m} |\Omega\rangle,
\end{equation}
where $i_k$ ($0\le i_k \le N$) denotes the number of bosons at the
$k^{th}$ site, and $m$ is the total number of sites. Note that we
restrict ourselves to the subspace of states with exactly $N$ bosons.

We construct a bosonic Hamiltonian ${\cal H}_{\rm bose}$, whose
ground-state energy determines the ground-state energy of the
following quantum spin glass model ${\cal H}$:
\begin{equation}
{\cal H}=\sum_{\langle i,j\rangle}\sum_{\mu,\nu=0}^3
c_{ij}^{\mu\nu}\sigma_i^{(\mu)}\otimes\sigma_j^{(\nu)},
\end{equation}
where $i, j$ label the site, $\sigma^{(0)}=\openone$ denotes the
identity and $\sigma^{(1)}=\sigma^x$, $\sigma^{(2)}=\sigma^y$, and
$\sigma^{(3)}=\sigma^z$ are the three Pauli matrices. The coefficients
$c_{ij}^{\mu\nu}$ are real but arbitrary. Oliveira and
Terhal~\cite{Terhal} showed that determining the ground-state energy
of ${\cal H}$ is QMA-hard, even if the interactions are restricted to
nearest neighbor sites $\langle i,j\rangle$ in the two dimensional
square lattice. By way of reduction, solving the ground-state energy
of ${\cal H}_{\rm bose}$ is also QMA-hard. To construct ${\cal H}_{\rm
  bose}$ , we use the Schwinger boson correspondence between qubit and
boson states (see, e.g., Ref.~\cite{Auerbach}) given by the following
map:
\begin{equation}
\sigma^x_i \leftrightarrow a_i^\dagger b_i + b_i^\dagger a_i, \ \sigma^y_i
\leftrightarrow i( b_i^\dagger a_i - a_i^\dagger b_i), \ \sigma^z_i
\leftrightarrow  a_i^\dagger a_i-b_i^\dagger b_i, \label{eqn:Schwinger}
\end{equation}
where the operators~$a_i,b_i$ correspond to distinct sites. (This is
similar to the mapping in the fermionic
case~\cite{LiuChristandlVerstraete07}.)  It is easy to verify
that the bosonic operators obey the commutation relations of the corresponding Pauli operators. We can regard the qubit at site $i$
as a single boson that can be in one of two different degrees of
freedom: $\ket{z_i}\leftrightarrow
(a_i^\dagger)^{z_i}(b_i^\dagger)^{1-z_i}\ket{\Omega}$ with $z_i\in
\{0,1\}$. This corresponds to the dual-rail encoding of a photonic
qubit in the Knill-Laflamme-Milburn linear-optics quantum computation
scheme~\cite{KLM}.  Hence, $N$ qubits can be represented by $N$ bosons
in $2N$ sites (or $N$ sites with each boson possessing two
distinct internal degrees of freedom):
\begin{equation}
\ket{z_1, \ldots, z_N} \leftrightarrow
   (a_1^\dagger)^{z_1} (b_1^\dagger)^{1-z_1} \cdots
   (a_N^\dagger)^{z_N} (b_N^\dagger)^{1-z_N} \ket{\Omega}.
\label{eqn:mapStates}
\end{equation}

Next, we consider the resulting bosonic Hamiltonian. As any two-local qubit
Hamiltonian can be written as a linear combination of terms with at most two
Pauli operators, the corresponding bosonic Hamiltonian can be written as a
combination of products of at most two annihilation and two creation
operators. To ensure that there be exactly one boson on the pair of sites
corresponding to~$i$, we add the following extra terms:
\begin{equation}
P_i\equiv(a_i^\dagger a_i +b_i^\dagger b_i-\openone)^2,
\end{equation}
which  commute with other terms in the Hamiltonian. The total bosonic
Hamiltonian is then
\begin{equation}
{\cal H}_{\rm bose}\equiv {\cal H}(a^\dagger,b^\dagger,a,b) + \sum_i c
\, P_i,
\end{equation}
which involves at most nearest-neighbor two-body interactions. By making the
weight $c$ of these projectors large enough, e.g., $\sum_{i,j,\mu,\nu} \left|
c_{ij}^{\mu\nu} \right|N(N-1)/2$, we guarantee that the ground state of the
full Hamiltonian has exactly one boson per site. Thus~${\mathcal H}_{\mathrm
bose}$ may be represented with at most a polynomially larger number of bits as
compared to~$\mathcal H$.  Thus, if one can compute the ground-state energy of
general bosonic Hamiltonians with at most two-body interactions, one can solve
general spin-1/2 two-local Hamiltonian problems. As solving the latter is
QMA-hard, solving the former is QMA-hard as well. This shows that
interacting boson problems are generally difficult.
\medskip

\noindent {\it QMA-hardness of bosonic $N$-representability
  problem.\/} We consider the number of bosons $N$ to be fixed, and
assume that the number of modes~$m$ that the bosons occupy is large enough,
i.e., $m \geq \delta N$ for some constant $\delta>0$. The number of different
ways ${\cal N}_m$ that $N$ identical bosons can occupy $m$ sites is ${\cal
N}_m=\binom{N+m-1}{N}$, which grows exponentially in $N$, i.e., ${\cal N}_m
\gtrsim(\delta\!+\!1)^N/\delta^N$ when $N$ is large. Given an $N$-boson state
$\rho^{(N)}$, the two-boson reduced density matrix is calculated by tracing
out all but two of the bosons: $\rho^{(2)} \equiv {\rm Tr}_{N-2} \rho^{(N)}$,
where $\rho^{(N)}$ is in general a mixture of states $\ket{\psi}$ of the
form~(\ref{eqn:psiN}) with exactly $N$ bosons. A precise definition of
$\rho^{(2)}$ is given via its matrix elements:
\begin{equation}
\rho^{(2)}_{ijkl} \equiv \frac{1}{N(N-1)}
\langle a_k^\dagger a_l^\dagger a_j a_i \rangle,
\end{equation}
where the bracket indicates the expectation value over the state
$\rho^{(N)}$.  Note that the one-boson reduced density matrix
$\rho^{(1)} \equiv {\rm Tr}_{N-1} \rho^{(N)}$,
defined via $\rho^{(1)}_{ik} \equiv \langle a_k^\dagger a_i
 \rangle/N$,
is completely determined once $\rho^{(2)}$ is known:
\begin{equation}
\label{eqn:rho1} \rho^{(1)}_{ik} = \sum_{l} \rho^{(2)}_{ilkl}.
\end{equation}

Informally, the bosonic $N$-representability problem (with $m$ sites) asks
whether there is an~$N$-boson state whose two-particle reduced density matrix
equals a given state~$\rho$. To deal with technical issues related to
precision, we are promised that when there is no~$N$-boson state consistent
with it, every two-particle reduced density matrix is ``far away''
from~$\rho$. Formally, we are given a two-boson density matrix $\rho$ of size
$[{m(m+1)}/{2}] \times [{m(m+1)}/{2}]$, and a real number $\beta \geq 1/{\rm
poly}(N)$, with all numbers specified with ${\rm poly}(N)$ bits of precision.
We would like to decide if:
 (``YES'' case) There exists an $N$-boson state $\sigma$
such that ${\rm Tr}_{N-2}(\sigma) = \rho$, or if (``NO'' case) For all $N$-boson states $\sigma$,  $\lVert
{\rm Tr}_{N-2}(\sigma) - \rho \rVert_1 \geq \beta$.

We show that the bosonic $N$-representability is QMA-hard under {\em
  Turing reductions\/}~\cite{Papadimitriou}. In other words, we show
that given an efficient algorithm for bosonic $N$-representability, we
can efficiently determine the ground state energy of ${\cal H}_{\rm
  bose}$, a QMA-hard problem as established above.  In the sequel, we
refer to an algorithm for $N$-representability as the ``membership
oracle''.

The first step is to write the two-particle interacting terms in
${\cal H}_{\rm bose}$ in terms of a complete orthonormal set, ${\cal
  \Q}$, of two-particle observables: $H_{\text{two-body}} \equiv \sum_{\Q
  \in {\cal \Q}} \gamma_{\Q}\, \Q$, where the number of elements ${l}
\equiv |{\cal \Q}| \sim O(m^4)$.  Note that ${\rm poly}(m)$ is ${\rm
  poly}(N)$, and so is ${\rm poly}(l)$. The observables~${\mathcal \Q}$
are constructed as in the fermionic case~\cite{LiuChristandlVerstraete07}.  We
define $a_I \equiv a_{i_2} a_{i_1}$, for all pairs $I = (i_1,i_2)$, $i_1\le
i_2$. (Note that in the case of fermions $i_1<i_2$.) We fix a total order
(denoted by $\prec$) on pairs of indices~$I$. The observables in ${\cal \Q}$
are defined as follows:
\begin{eqnarray}
X_{IJ} &\equiv& \frac{1}{\sqrt{n_I n_J}}( a_I^\dagger
a_J + a_J^\dagger a_I), \ \ \mbox{for} \, I \prec J,\\
Y_{IJ}& \equiv& \frac{-i}{\sqrt{n_I n_J}}(
a_I^\dagger a_J - a_J^\dagger a_I), \ \ \mbox{for} \, I \prec J,\\
Z_I &\equiv& \frac{1}{n_I}a_I^\dagger a_I, \ \ \mbox{for} \, I\prec L,
\end{eqnarray}
where the factor $n_I=1$ if $i_1\ne i_2$, $n_I=2$ if $i_1=i_2$, and
$L$ denotes the last pair in the ordering. These operators are
Hermitian and the $X_{IJ}$ and $Y_{IJ}$ have expectation values in the
interval $[-1,1]$ and the $Z_I$ have expectation values in $[0,1]$ for any
two-particle state. In the two-particle Hilbert space, they are
orthogonal to each other under the trace operator, e.g., ${\rm
  Tr}(X_{IJ} Z_K)=0$ for all~$I,J,K$. They also form a basis
for the two-boson density matrices, i.e., any such matrix~$\rho^{(2)}$ may be
written as
\begin{eqnarray}
\label{eqn:rho2}
\rho^{(2)}& = &Z_L +\sum_{I\prec L} \alpha_{Z_I}(Z_I-Z_L)\nonumber \\
  && \mbox{} + \frac{1}{2} \sum_{I\prec J}
     \left( \alpha_{(X_{IJ})} X_{IJ}+\alpha_{(Y_{IJ})}Y_{IJ} \right),
\end{eqnarray}
where $\alpha_\Q = {\rm Tr}(\Q\rho^{(2)})$ for all $\Q \in {\cal Q}$. Using
Eq.~(\ref{eqn:rho1}) we see that the expectation value $\langle a_i^\dagger
a_k\rangle$ of the one-body terms, can be expressed as linear combination of
the $\alpha_{\Q}$.  Thus we have ${\rm Tr}(H_{\rm
  bose}\rho^{(N)}) = \sum_{\Q \in {\cal \Q}} \gamma_{\Q}' \, \alpha_{\Q}$,
where $\gamma_{\Q}'$ includes the contribution from both one-body and two-body
terms, and $\alpha_{\Q}$ are the coefficients of $\rho^{(2)} \equiv {\rm
Tr}_{N-2}\big(\rho^{(N)}\big)$. Finding the ground-state energy is equivalent
to minimizing the linear function $f(\vec{\alpha}) = \sum_{\Q \in {\cal Q}}
\gamma_{\Q}'\, \alpha_{\Q}$ subject to the constraint that $\vec{\alpha} \in
K_N$, where $K_N$ denotes the convex set of all $\vec{\alpha}$ such that the
corresponding state $\rho^{(2)}$ is $N$-representable.

The above minimization of energy, subject to the convex constraint that $\rho$
is $N$-representable, belongs to a class of convex optimization problems which
can be solved using the shallow-cut ellipsoid algorithm~\cite{book,LiuThesis}
with the aid of an $N$-representability membership oracle.  If~$K_N$ is
contained in a ball of radius $R$ centred at the origin, and it contains a
ball of radius $r$, the run time is ${\rm poly}\big(\log(R/r)\big)$ and the
error in the solution due to computation with finite precision is
$1/{\rm poly}(R/r)$. The algorithm will be efficient and of polynomially
bounded error, if $R/r$ is at most ${\rm poly}(l)$.

{}From the discussion leading to Eq.~(\ref{eqn:rho2}) it follows that $K_N$ is
contained in a ball of radius $R=\sqrt{l}$ centered at the origin. However,
the method used by Liu et al.~\cite{LiuChristandlVerstraete07} of regarding
$a_j$ as creation operator for a hole in site $j$ cannot be employed to
suitably bound $r$ from below. Instead, we explicitly construct a set of
$N$-boson states such that the convex hull of the corresponding vectors
$\{{\alpha}_{\Q}\}$ contains a ball of radius $1/{\rm
  poly}(l)$. Relying on the property that
bosons can occupy the same site, for each~$Q$ we construct states so as
to maximize or minimize $\alpha_Q$. The resulting points have
the property that for any coordinate axis, there exist at least two
points at constant distance along that axis. As a consequence, we show
that their convex hull (which is contained in $K_N$) contains a ball
with radius $r\ge 1/{\rm poly}(l)$, centered at the center of mass of
the points.

An algorithm for bosonic $N$-representability thus enables efficient
calculation of the ground-state energy of the interacting boson
Hamiltonian ${\cal H}_{\rm bose}$. As a consequence,
$N$-representability is QMA-hard. This may come as a surprise, as bosonic
problems are generally regarded as being easier than the corresponding
fermionic problems~\cite{TroyerWiese}. \medskip

\noindent {\it QMA-completeness.\/} Is the bosonic
$N$-representability inside QMA or even harder? We show that the
bosonic $N$-representability problem is indeed inside QMA, implying
that the problem is QMA-complete. To establish this, we construct a
QMA proof system, i.e., describe a witness state $\tau$ (over
polynomially many qubits) for the $N$-representability of a given
two-boson density matrix~$\rho$, and a polynomial-time quantum
algorithm $V$ (the ``verifier'') that expects such a pair~$\rho,\tau$
as input.  The verifier determines probabilistically whether a given
two-boson density matrix~$\rho$ is $N$-representable, or is far from
being $N$-representable.  In the ``YES'' case, $V$ outputs ``YES''
with probability $p_1 = 1$, and in the ``NO'' case, $V$ outputs ``NO''
with probability~$\geq p_0 = 1-1/{\rm poly}(N)$.  The gap~$p_1 - p_0$
can be amplified to~$1 - \e^{-{\mathrm
    {poly}}(N)}$~\cite{Kitaev,MW_NWZ}.

For the witness state, we represent an $N$-boson state $\sigma$ using
$m$ qu$d$its ($d$-dimensional quantum systems), via the following
correspondence, a.k.a.\ Holstein-Primakoff bosons (see, e.g.,
Ref.~\cite{Auerbach}):
\begin{equation}
  a_i \leftrightarrow A_i \equiv \frac{1}{\sqrt{s+S_i^z}}S_i^{+}, \ \
  a^{\dagger}_i \leftrightarrow A_i^\dagger \equiv
                                \frac{1}{\sqrt{s+S_i^z+1}}S_i^{-},
\end{equation}
where $S^\pm_i$ are the raising/lowering operators for $i$'th spin, and $2s\ge
N$.  The spin operators above satisfy the bosonic commutativity relations
provided the total spin magnitude is $s$. The boson number states $\ket{n} \in
\{\ket{0},\ket{1},...,\ket{2s}\}$ at one site correspond to the spin states
$\ket{s_n} \in \{\ket{s},\ket{s-1},...,\ket{-s}\}$, and~$d = 2s+1$.  A bosonic
observable $O = a_i^\dagger a_j^\dagger a_l a_k + a_k^\dagger a_l^\dagger a_j
a_i$ is transformed into $\tilde{O} = A_i^\dagger A_j^\dagger A_l A_k +
A_k^\dagger A_l^\dagger A_j A_i$, which is a tensor product of at most four
single-qu$d$it observables, in contrast to the involvement of non-local
Jordan-Wigner string operators in the fermionic
case~\cite{LiuChristandlVerstraete07}.

The expectation value $\langle \tilde{O} \rangle$ can be estimated
efficiently. One method is to explicitly diagonalize the
observable~$\tilde{O}$ as~$\sum_i \lambda_i \ketbra{\theta_i}$, and measure
the given qu$d$it representation~$\tilde{\sigma}$ of the bosonic
state~$\sigma$ in the basis~$\{ \ket{\theta_i} \}$. Repeating the measurement
on polynomially many copies of~$\tilde{\sigma}$, we can estimate~$\langle
\tilde{O} \rangle$. We note that a quantum circuit using qu$d$its with
$d=2s+1$ can be implemented efficiently by an equivalent circuit using qubits.

The witness $\tau$ consists of polynomially many blocks, where each
block has $m$ qu$d$its that represent a state $\tilde{\sigma}$ that is
claimed to be an $N$-boson state with~${\mathrm
  Tr}_{N-2}(\tilde{\sigma}) = \rho$.  The verifier $V$ measures, on
each block, the observable $\sum_k {A^\dagger_k A_k}=ms-\sum_k S_k^z$ to check
whether the particle number is $N$. If it is not $N$, $V$ outputs ``NO''. This
measurement projects each block onto the space of fixed particle number
states. If the particle number is $N$, $V$ continues to perform measurements
for a suitable set of observables (e.g., those corresponding to ${\cal \Q}$)
using the projected states. It compares whether the outcomes match the
expectation values specified by $\rho$, to check for consistency. It outputs
``YES'' if the errors are less than $\beta/{\rm poly}(N)$ (for a suitable
polynomial), otherwise outputs ``NO''.  When $\rho$ is $N$-representable, the
prover supplies polynomially many copies of the state $\sigma$ such that ${\rm
Tr}_{N-2}(\sigma)=\rho$, the verifier always answers ``YES'' (i.e., $p_1=1$).
When $\rho$ is not $N$-representable, the prover can cheat by entangling
different blocks of qudits. Using a Markov argument, as was first employed by
Aharonov and Regev~\cite{Aharonov} and later by Liu~\cite{LiuThesis}, one can
show that the verifier will still output ``NO'' with probability $\ge
\beta/{\rm poly}(N)$.
  Thus $N$-representability is in
QMA~\cite{QMA1}.

\noindent {\it  Further consequences.\/}
Now that we have shown the QMA-completeness of $N$-representability
for bosons, we may follow the same argument as by Liu, Christandl, and
Verstraete~\cite{LiuChristandlVerstraete07} and reach the conclusion
that {\it pure-state\/} bosonic $N$-representability is in
QMA($k$). This is achieved because the essential point is to verify
the purity of the certificate. Next, consider the problem of deciding
bosonic $N$-representability of the two-boson density matrix when only
the diagonal elements $D_{ij}\equiv\langle a_i^\dagger a_j^\dagger a_j
a_i\rangle$ are specified. If one considers the case $m=2N$ and the
mapping by the Schwinger representation~(\ref{eqn:Schwinger}), one
finds that as in the fermionic case, the solution enables one to solve
the ground-state energy of local spin Hamiltonians which only contain
commuting $\sigma^z$ operators. The latter corresponds to a classical
spin-glass problem, and is known to be NP-hard~\cite{Barahona}. Thus
the problem of deciding $N$-representability given $\{D_{ij}\}$ is
also NP-hard.  \medskip

\noindent {\it Acknowledgments\/}. TCW thanks Zhengfeng Ji for useful
discussions. This work was supported by ARO/NSA (USA), CIFAR, IQC, MITACS,
NSERC, an Ontario ERA, ORF, and QuantumWorks.


\begin{thebibliography}{99}
\bibitem{QMA1} What we have shown here is that
$N$-representability problem is inside
${\rm QMA_{1}}$, i.e., in QMA with one-sided error. Since we have also
shown that this problem is QMA-hard, we conclude its QMA-completeness.
It is not known whether every problem in QMA has a proof system with
one-sided error.
See the dicussion by S. Aaronson, Quantum Inf. Comput. {\bf 9}, 81 (2006).
\end{thebibliography}
\end{document}